\documentstyle[12pt]{article}
\begin{document}
\thispagestyle{empty}
\vspace*{1cm}
\begin{center}
{\large Spin dynamics
in the generalized\\[0.2cm] ferromagnetic Kondo model for manganites.}\\[1.cm]
\end{center}
\begin{center}
{N.B.Perkins and N.M.Plakida}\\
\end{center}
\begin{center}
{\it Joint Institute for Nuclear Research, Dubna 141980, Russia }
\end{center}
%\vspace*{2cm}

Dynamical spin susceptibility is calculated for the generalized
ferromagnetic Kondo model  which describes itinerant
$e_{g}$ electrons interacting with localized $t_{2g}$ electrons
with antiferromagnetic coupling. The calculations done in the mean field
approximation show that the spin-wave spectrum of the system in ferromagnetic
state has two branches,  acoustic and optic ones. Self-energy corrections
to the spectrum are considered and the acoustic spin-wave damping is
evaluated.

\newpage
\section{Introduction}

	Manganites of the perovskite structure of the form $R_{1-x}B_xMnO_3$ (where
$R$ are trivalent rare-earth and $B$ are divalent alcaline ions, respectively)
 and related compounds that present
the phenomenon of "colossal" magnetoresistance (CMR) have recently
 attracted much attention both from the basic point of view and due to
 their potential application \cite{ram}, \cite{mil}. The large magnetoresistance
 occurs close to
 the metal-insulator and the paramagnetic-ferromagnetic transitions where the
 interplay of transport, magnetic and structural properties is of the great
 importance (see \cite{khom}).
 The key elements of the manganese oxides are $Mn$ ions. In the
 parent compound
 $La Mn O_3$ the electronic configuration of $Mn^{+3}$  is $(t^3_{2g} e_g)$.
In this configuration due to a strong intra-orbital Hund's coupling  $t_{2g}^3$
 electrons go into tightly  bound $d_{xy},~d_{yz},~d_{xz}$ core
 states and make up an electrically inert
  core Heisenberg spins $S$ of magnitude $3/2$. The $t_{2g}^3$
 configuration is very stable and remains localized over the entire
 range of doping.

 In the undoped case, with one $e_g$ electron
 per $Mn$ ion,
two $e_g$ orbitals, $d_{x^2-y^2}$ and $d_{3z^2-r^2}$ types, are splitted due
to the Jahn-Teller effect. At low temperature  $ e_g$ electrons occupy
$d_{3x^2-r^2}$ and $d_{3y^2-r^2}$ ordered alternately in $ab$ plane with
their spins aligned to the core spin by interorbital Hund's coupling.
Due to the Goodenough-Kanamori rules \cite{good} it
results in the A-type antiferromagnet (AFM) ground state (AFM vector
Q=(0,~0,~0.5))
with spin $S=2$ for $LaMnO_3$. Upon doping
 with holes by substituting $La$ with $Sr$ or any other divalent ions system
 becomes ferromagnetic (FM) and conducting. The hopping between two
 $Mn$ sites is maximal when
 the core spins are parallel and minimum when they are antiparallel.That
 results in effective ferromagnetic exchange between the nearest neighbor
 core spins    and thus leads to the FM metallic ground state of doped compounds.
 This behavior is qualitatively well described within the framework of the
double exchange  (DE) mechanism (see \cite{zener},\cite{and},\cite{degen}). At higher hole
concentration, $x\geq 0.5$, a charge ordering for holes is observed, and at
$x=1$ an insulating G-type AFM state (with $Q=(0.5,~0.5,~0.5)$) takes place for
$CaMnO_3$ compound. Therefore to describe the experimentally obtained phase
diagram ( see, for example, \cite{shif}) one should take into account both the
Heisenberg type of AFM exchange between the core $t_{2g}$ electrons and
the strong Hund coupling between $t_{2g}$ and $e_{g}$ electrons (see, e.g.
\cite{ino,ish}). These competing interactions could be responsible for a
coexistence of AFM and FM states observed recently in neutron scattering
experiments in $(La_{0.25}Pr_{0.75})_{0.7}Ca_{0.3}MnO_3$ \cite{balag} and
in the bilayer manganite $La_{1.2}Sr_{1.8}Mn_{2}O_{7}$ \cite{per}.
Also a crossover from an ideal isotropic FM spin-wave behaviour at low
temperature to a diffusive spin propagation observed  in
$La_{0.7}Ca_{0.3}MnO_{3}$ \cite{lynn} could be explained if one takes into
 account both the localized $t_{2g}$ spin ($S=3/2$) and the itinerant $e_g$
 spin ($\sigma=1/2$) subsystems.

In the present paper we study the spin dynamics in manganites within the
generalized ferromegnetic Kondo model (FKM)
allowing for the AFM exchange interaction between $t_{2g}$ spins.
Unlike to the DE model (see, \cite{fur}), where
$J_H/t\rightarrow\infty$ is considered and the system is treated as
 perfectly spin polarized with $S=2$, in our work both the fluctuation
 of the localized  and itinerant spins are taking into account.
However,
we ignore in the present calculations a possible orbital ordering which is
very important in explaining different types of AFM ordering in the insulating
phases \cite{ish,fei} but plays less essential role in the FM state
considered here. To take into account strong Coulomb interaction between
$e_g$ electrons which excludes  the double occupancy of $e_g$ electrons
for a lattice site we employ the Hubbard operator technique. The spectrum of spin
waves in the FM state is calculated by employing  equations of motion for
the matrix Green function (GF) for the localized and itinerant spins.
In the next Section the model and general formalism for the GF are presented.
The spin-wave spectrum in a generalized mean field approximation (MFA) is
calculated in Sec. 3 and  self-energy corrections and spin-wave
damping are evaluated in Sec. 4.

\section{The model}

We consider an effective Hamiltonian of the generalized FM Kondo model
which can be written  in the following form \cite{ino}:
\begin{equation}
H=-\sum_{i,j,\sigma}t_{ij}X_{i}^{\sigma 0}X_{j}^{0\sigma}-\frac{J_H}{2S}
\sum_{i}{\bf S}_i{\bf{\sigma}}_i
+\frac{1}{2}\sum_{i,j}J_{ij}{\bf S}_{i}{\bf S}_{j}~.
\label{m1}
\end{equation}
The first term  of Eq. (\ref{m1}) describes an electron
hopping between Mn-ions where
$X_{i}^{\sigma 0}$ is the creation operator of an electron with spin
$\sigma$  in one of the $e_g$ orbitals. Here  we neglect
orbital degeneracy of $e_g$ electrons and introduce orbital independent
hopping parameter $t_{ij}$ with $t_{ij}=t$ for the nearest neighbors.
The second term describes the ferromagnetic Hund coupling
($J_H>0$) between $e_g$ and $t_{2g}$ spins
where ${\bf S}_i$ refers to the localized Mn core spin $S=3/2$.
The third term
describes the antiferromagnetic coupling of localized
spins between the nearest neighbor sites. In real materials
the coupling of core spins is not the same in different directions
and should be  described in the matrix form, but for  simplicity we
 are analyzing the isotropic case ($J_{ij}=J$).
We exclude the doubly occupied $e_{g}$ state from the effective
Hamiltonian by using the Hubbard operator representation
because the electron-electron interaction has the largest
energy scale (intra-atomic Coulomb interaction in the $e_g$ orbitals) and
can be estimated as $7-8~$eV while $J_H \sim 1$~eV.  Due to large Hund
energy we neglect superexchange interaction between $e_{g}$ electrons
of the order of $t^{2}/U$~\cite{ish}.
The conduction bandwidth is smaller than the Hund
coupling energy and from density-functional studies
 can be estimated as  $t \simeq 0.15$~eV~\cite{sap}.

The HO's in Eq. (\ref{m1}) are defined as
$X_{i}^{\alpha\beta}=|i,\alpha\rangle\langle i,\beta|$
for three possible states at the lattice site $i$:
$|i,\alpha\rangle=|i,0\rangle,\;\;\;|i,\sigma\rangle$ for an empty site
and for a singly occupied site with spin
$\sigma = (\uparrow, \downarrow) = ( +, - ) \;$.
The completeness relation for the HO's reads as
\begin{equation}
X_{i}^{00}+\sum_{\sigma}X_{i}^{\sigma\sigma}=1.
\label{m4}
\end{equation}
For itinerant electrons the spin and density operators
in Eq. (\ref{m1}) are expressed by
HO's as
\begin{equation}
\sigma_{i}^{+}=X_{i}^{\uparrow\downarrow},\;
\sigma_{i}^{-}=X_{i}^{\downarrow\uparrow},\;
\sigma_{i}^{z}=\frac{1}{2}( X_{i}^{\uparrow\uparrow}-
X_i^{\downarrow\downarrow}),\quad
n_{i}=X_{i}^{\uparrow\uparrow}+X_i^{\downarrow\downarrow}.
\label{m5}
\end{equation}
 The HO's obey the following
commutation relations
\begin{equation}
\left[X_{i}^{\alpha\beta}, X_{j}^{\gamma\delta}\right]_{\pm}=
\delta_{ij}\left(\delta_{\beta\gamma}X_{i}^{\alpha\delta}\pm
\delta_{\delta\alpha}X_{i}^{\gamma\beta}\right).
\label{m7}
\end{equation}
In Eq.(\ref{m7}) the upper sign stands for the case when both
HO's are Fermi-like ones (as, e. g., $X_{i}^{0\sigma}$).
The spin and density operators (\ref{m5}) are Bose-like and for them
the lower sign in Eq.(\ref{m7}) should be taken.

It is assumed that the core spin  operators  $ S^{\alpha}_{i}$
obey the standard commutation relations, e.g.,
\begin{eqnarray}
\left[ S_i^{+}, S_j^{-}\right]=2\delta_{i,j} \; S_j^z~.
\label{m8}
\end{eqnarray}

To treat the fluctuations of localized and itinerant spins
at the same level of approximation we introduce the
 dynamic spin susceptibility (DSS) of the system in
the matrix form
\begin{equation}
\chi (q, \omega) =\left(
\begin{array}{cc}
\chi_{11} & \chi_{12}\\
\chi_{21} & \chi_{22}\\
\end{array}\right)
=\langle\langle A_q|A_q^+\rangle\rangle_{\omega}~,
\label{g1}
\end{equation}
where
$$
A_q=\left(
\begin{array}{c}
\sigma^+_q\\
S^+_q\\
\end{array}
\right),~~
A_q^+=\left(
\sigma^-_q ~~ S^-_q \right)~.
$$
Here
\begin{equation}
\langle\langle A_q|A_q^+\rangle\rangle_{\omega}= -\imath
\int_0^{\infty} dt e^{-\imath \omega t} \frac{1}{N} \sum_{q}
e^{-\imath q (l-m)}
\langle [ A_{l}(t), A_{m}^{+}]\rangle
\label{m22}
\end{equation}
denotes the Fourier transformed two-time retarded
commutator Green function (GF) ~\cite{zub,tse}.
The diagonal elements $\chi_{11}(q, \omega)$ and $\chi_{22}(q, \omega)$
 stands for the
 itinerant and core spin GF, respectively, while the nondiagonal
elements $\chi_{12}(q,\omega)$ and $\chi_{21}(q,\omega)$ define the
crosscorrelations
between the  two spin subsystems. The GF (\ref{g1}) obeys the following
equation of motion
\begin{eqnarray}
\omega\langle\langle A_q|A^{+}_q\rangle\rangle_{\omega}=
\langle[A_q, A_q^{+}]\rangle+
\langle\langle \imath {\dot A}_{q}|A^{+}_q\rangle\rangle_{\omega}~,
\nonumber\\
\omega\langle\langle \imath {\dot A}_{q}| A^{+}_q\rangle\rangle_{\omega}=
\langle[\imath {\dot A}_{q}, A_q^{+}]\rangle+
\langle\langle \imath {\dot A}{_q}|-\imath {\dot A}^{+}_{q}\rangle\rangle_{\omega}~.
\label{m23}
\end{eqnarray}
These equations (\ref{m23}) could be easily combined in a more convenient
form of the equation of motion \cite{tse}:
$$
\omega\langle\langle A_q | A^{+}_q\rangle\rangle_{\omega}=
\langle [A_q,A_q^{+}]\rangle
$$
\begin{equation}
+ \left( \langle[\imath{\dot A}_{q}, A_q^+]\rangle +
\langle\langle\imath{\dot A}_{q}|-\imath{\dot A}_{q}^{+}
\rangle\rangle_{\omega}^{irr} \right)
\cdot\frac{1}{\langle [A_q , A_q^+]\rangle}\cdot
\langle\langle A_q|A_q^+\rangle\rangle_{\omega}~,
\label{g2}
\end{equation}
where the current is defined as
$\imath{\dot A}=\imath dA/dt=[A,H]$ and in the matrix form
can be given by the following expression:
\begin{equation}
\imath{\dot A}_{q}=\left(
\begin{array}{c}
\imath{\dot \sigma}^{+}_{q}\\
\imath{\dot S}^{+}_{q}\\
\end{array}
\right)~,
\label{g21}
\end{equation}
and
$$
\langle\langle\imath{\dot A}_{q}|-\imath{\dot A}_{q}^{+}
\rangle\rangle_{\omega}^{irr}
=
\langle\langle\imath{\dot A}_{q}|-\imath{\dot A}_{q}^{+}
\rangle\rangle_{\omega}
$$
\begin{equation}
  - \langle\langle\imath\dot{A_q}|A_q^+\rangle\rangle_{\omega}
\langle\langle A_q| A_q^+\rangle\rangle_{\omega}^{-1}
\langle\langle A_q|-\imath\dot{A_q^+}
\rangle\rangle_{\omega}
\label{g20}
\end{equation}
is the irreducible part of the higher order GF.

We can rewrite (\ref{g2}) in the Dyson form
\begin{equation}
\chi_{q}(\omega)= [\omega\hat{\tau}_{0}-{\tilde\Omega}_{q} -
{\tilde\Pi}(q,\omega)]^{-1} \cdot I ,
\label{g3}
\end{equation}
where $\hat{\tau}_{0}$ is the unity matrix and
\begin{equation}
I=\langle [A_q, A_q^+]\rangle=
\left(
\begin{array}{cc}
\langle[\sigma_q^+, \sigma_q^-]\rangle &
\langle[\sigma_q^+, S_q^-]\rangle \\
\langle[S_q^+, \sigma_q^-]\rangle &
\langle[S_q^+, S_q^-]\rangle \\
\end{array}
\right)=
\left(
\begin{array}{cc}
2\langle\sigma^z \rangle & 0 \\
0 & 2\langle S^z \rangle\\
\end{array}
\right)
\label{g4}
\end{equation}
where
$\langle \sigma^z \rangle = \langle \sigma_{l}^z \rangle $
and
$\langle S^z \rangle = \langle S_{l}^z \rangle $ .

The matrix ${\tilde\Omega}_{q}=\Omega_{q}\; I^{-1}$ describes
the  mean field (MF)
energy spectrum and  ${\tilde\Pi}(q,\omega)=\Pi(q,\omega) \; I^{-1}$
is the self-energy matrix. They are given by
\begin{equation}
\Omega_{q}=\langle[\imath \dot{A_q}, A_q^+]\rangle=
\left(
\begin{array}{cc}
\langle[\imath\dot{\sigma_q^+}, \sigma_q^-]\rangle &
\langle[\imath\dot{\sigma_q^+}, S_q^-]\rangle \\
\langle[\imath\dot{S_q^+}, \sigma_q^-]\rangle &
\langle[\imath\dot{S_q^+}, S_q^-]\rangle \\
\end{array}
\right)  \; ,
\label{g5}
\end{equation}

\begin{equation}
\Pi(q,\omega) =\langle\langle\imath \dot{A_q}|-\imath\dot{ A_q^+}\rangle\rangle^{irr}=
\left(
\begin{array}{cc}
\langle\langle\imath\dot{\sigma_q^+}| \imath\dot{\sigma_q^-}\rangle\rangle^{irr} &
\langle\langle\imath\dot{\sigma_q^+}| \imath\dot{S_q^-}\rangle\rangle^{irr} \\
\langle\langle\imath\dot{S_q^+}| \imath\dot{\sigma_q^-}\rangle\rangle^{irr} &
\langle\langle\imath\dot{S_q^+}| \imath\dot{S_q^-}\rangle\rangle^{irr} \\
\end{array}
\right)  \; ,
\label{g6}
\end{equation}
with
\begin{equation}
\imath \dot {\sigma^+_l}
=\sum_i t_{il}(X_i^{\uparrow0} X^{0\downarrow}_l -
X_l^{\uparrow0} X_i^{0\downarrow})
-\frac{J_H}{2S}(S_{l}^{+}\sigma_{l}^{z}-S_{l}^{z}\sigma_{l}^{+})~,
\label{g22}
\end{equation}

\begin{equation}
\imath \dot {S^+_l}
=-\frac{J_H}{2S}(S_{l}^{z}\sigma_{l}^{+}-
S_{l}^{+}\sigma_{l}^{z})-\sum_i J_{il}(S_i^z S^{+}_l-S_{l}^{z}S_{i}^{+})~.
\label{g23}
\end{equation}

\section{Mean field approximation}
Let us now examine the spectrum and  DSS in mean field approximation (MFA).
The spin-wave dispersion is
 determined by the following equation
\begin{equation}
\det~\left(\omega \hat{\tau}_{0} - {\tilde\Omega}_{q}\right) =0 ~.
\label{g8}
\end{equation}
From (\ref{g5}) we obtain for the matrix elements of ${\tilde\Omega}_q$
\begin{equation}
{\tilde\Omega}_q =
\left(
\begin{array}{cc}
[d +a(1-\gamma_q)]/2\langle\sigma^{z}\rangle
& -d/2\langle S^{z}\rangle \\
-d/2\langle\sigma^{z}\rangle  &
[d -b(1-\gamma_q)]/2\langle S^{z}\rangle \\
\end{array}
\right)~,
\label{g10}
\end{equation}
where we are using the following notation:
\begin{equation}
d= \frac{J_H}{2S}(2\langle\sigma_l^z S_{l}^z\rangle+
\langle\sigma_l^+ S_{l}^-\rangle )~,
\end{equation}
\begin{equation}
a=zt(n_1^{\uparrow}+n_1^{\downarrow})~,~~b=zJN_1 ,
\label{g32}
\end{equation}
with  $t_q=zt\gamma_q, \; \gamma_q= (2/z) (\cos q_x+\cos q_y+\cos q_z)$,
where  $z=6$ for the simple three-dimensional
cubic lattice with nearest-neighbor hopping $t$.
 In (~\ref{g32}) the nearest neighbor particle-hole
 and spin correlation functions
are defined as follows
\begin{eqnarray}
n_1^{\sigma}=\frac{1}{N}\sum_k \gamma_k n_k^{\sigma} ~,~~n_k^{\sigma}=
 \langle X_k^{\sigma 0} X_k^{0\sigma}\rangle \nonumber\\
N_1=\frac{1}{N}\sum_k \gamma_k N_k~,~~N_k=
 2\langle S_k^zS_{-k}^z \rangle + \langle S_k^+S_{k}^- \rangle ~.
\label{g33}
\end{eqnarray}
The equation (\ref{g8}) has two solutions describing two branches of
spin wave excitations:
\begin{eqnarray}
E^{1(2)}_{q}=\frac{1}{2}\left[
{\tilde\Omega}_q^{11}+{\tilde\Omega}_q^{22} \mp  \sqrt{\left(
{\tilde\Omega}_q^{11}-{\tilde\Omega}_q^{22}\right)^{2}+4{\tilde\Omega}_q^{12}
{\tilde\Omega}_q^{21}}\right ].
\label{E12}
\end{eqnarray}
For the model calculation we can expand this equation
at $q\rightarrow 0$ and for the finite value of $d$ we obtain
\begin{eqnarray}
E_q^{(1)}\simeq D_1q^2        ,
\nonumber \\
E_q^{(2)}\simeq \Delta+D_2 q^2  ~,
\label{g13}
\end{eqnarray}
where $E^{(1)}_q$ corresponds to the gapless (acoustic)
spin-wave excitation with the stiffness $D_1$ given by
\begin{equation}
D_1=\frac{a-b}{12(\langle S^z \rangle +\langle \sigma ^z\rangle)}~,
\label{d1}
\end{equation}
 and  $E^{(2)}_q$ describes the optic mode of
the spin fluctuations
 with the gap $\Delta$ and the effective stiffness $D_2$ determined by
 the following expressions:
\begin{eqnarray}
\Delta=d\frac{\langle S^z \rangle +\langle \sigma ^z\rangle}
{2\langle S^z\rangle\langle \sigma ^z\rangle},~\nonumber\\
D_2=\frac{a\langle S^z \rangle^2 -b\langle \sigma ^z\rangle^2}
{12 \langle S^z\rangle\langle \sigma ^z\rangle
(\langle S^z\rangle+\langle \sigma ^z\rangle)}                    ~.
\label{d2}
\end{eqnarray}
The ferromagnetic acoustic spin-wave becomes  unstable when the
stiffness $D_1 \rightarrow 0$ or $ a-b =0 $ in Eq.~(\ref{d1}).
It may happen for small concentration of itinerant electrons,
$n \le n_{c} \simeq 2SJ/t \simeq 0.3$. The self-energy corrections
considered below (see Eq.~(\ref{ddd})) even increase the critical
value $n_c$.

The spectrum of spin fluctuations in MFA are given by the spectral
functions
\begin{equation}
 B_{\alpha\beta}^{MF} (q, \omega) = -\frac{1}{\pi}{\rm Im}
 \chi_{\alpha\beta }^{MF}(q,\omega+\imath \varepsilon)
\label{chi}
\end{equation}
for the spin susceptibility
\begin{eqnarray}
 B_{11}^{MF}(q, \omega)= 2 \langle \sigma ^z\rangle
\left( \frac{ {\tilde\Omega}^{22}_q - E^{(1)}_q }{E^{(2)}_q-E_q^{(1)}}
\delta(\omega-E^{(1)}_q) +
\frac{ { E^{(2)}_q - \tilde\Omega}^{22}_q }{E^{(2)}_q-E_q^{(1)}}
\delta (\omega-E_q^{(2)}) \right) \; ,
\label{chi11}
\end{eqnarray}
\begin{eqnarray}
 B_{22}^{MF}(q, \omega)= 2 \langle S^z\rangle
\left( \frac{  {\tilde\Omega}^{11}_q - E^{(1)}_q }{E^{(2)}_q-E_q^{(1)}}
\delta(\omega-E^{(1)}_q) +
\frac{ { E^{(2)}_q - \tilde\Omega}^{11}_q }{E^{(2)}_q-E_q^{(1)}}
\delta (\omega-E_q^{(2)}) \right) \; ,
\label{chi22}
\end{eqnarray}
\begin{eqnarray}
 B_{12}^{MF}(q, \omega)=  B_{21}^{MF}(q, \omega)=
 \frac{d}{E^{(2)}_q-E_q^{(1)}}
\left(\delta(\omega-E^{(1)}_q) - \delta (\omega-E_q^{(2)})\right) \; .
\label{chi12}
\end{eqnarray}
The spectral functions (\ref{chi11}) - (\ref{chi12}) obey the following
sum rules:
$$
  \int_{-\infty}^{+\infty} d\omega  B_{\alpha\beta}^{MF} (q, \omega) =
  I_{\alpha\beta }\; ,
$$
\begin{equation}
  \int_{-\infty}^{+\infty}\omega d\omega  B_{\alpha\beta}^{MF} (q, \omega) =
 \Omega_{q}^{\alpha\beta }=
 \langle[\imath \dot{A_q}, A_q^+]\rangle_{\alpha\beta }  \; ,
\end{equation}
where the matrices $I_{\alpha\beta }$ and $ \Omega_{q}^{\alpha\beta } $
are given by the Eqs.(\ref{g4}), (\ref{g5}).

\section{Self-energy corrections}

The next step is to consider  the self-energy corrections to the
MF spectrum.
Taking into account the self-energy  corrections  the equation
for the spectrum  transforms into the following form:

\begin{equation}
\det~\left(\omega \hat{\tau}_{0} -
{\tilde\Omega}_{q}-{\tilde \Pi}(q,\omega)\right) =0.
\label{g109}
\end{equation}
First we compute the self-energy matrix elements by using mode-coupling
approximation in terms of the dressed particle-hole and spin fluctuations
(see, e. g., G\"{o}tze et al.,~\cite{got}).
 This scheme is essentially equivalent to
the self-consistent Born approximation in which the vertex corrections
are neglected. The proposed scheme is defined by the following decoupling of the
time-dependent correlation functions:
\begin{equation}
\langle X_{m}^{-0}(t)X_{i}^{0+}(t)X_{j}^{+0}X_{l}^{0-}\rangle\simeq
\langle X_{m}^{-0}(t)X_{l}^{0-}\rangle
\langle X_{i}^{0+}(t)X_{j}^{+0}\rangle,
\label{g24}
\end{equation}
\begin{equation}
\langle \sigma_{i}^{z}(t)S_{m}^{-}(t)\sigma_{j}^{z}S_{l}^{+}\rangle\simeq
\langle \sigma_{i}^{z}(t)\sigma_{j}^{z}\rangle
\langle S_{m}^{-}(t)S_{l}^{+}\rangle \; ,
\label{g24a}
\end{equation}
\begin{equation}
\langle S_{i}^{z}(t)S_{m}^{-}(t)S_{j}^{z}S_{l}^{+}\rangle\simeq
\langle S_{i}^{z}(t)S_{j}^{z}\rangle
\langle S_{m}^{-}(t)S_{l}^{+}\rangle \; .
\label{g25}
\end{equation}

The self-energy matrix elements are obtained by using
the above defined decoupling scheme (\ref{g24}) and (\ref{g25})
with the  spectral representation for the  GF.
The diagonal elements involve two contributions:
\begin{equation}
\Pi_{11(22)}(q,\omega)=
\Pi_{11(22)}^{(1)}(q,\omega)
+
\Pi_{11(22)}^{(2)}(q,\omega)\; .
\label{g70}
\end{equation}
The first one describes  fluctuations of
the internal degrees of freedom of the given spin subsystem.
While the second one stems from the Hund's term and describes the
coupling between itinerant and core spins.

For the itinerant spins the first term in Eq. (\ref{g70}) is due to
 decay of  spin fluctuations
into particle-hole pair excitations and reads as
$$
\Pi_{11}^{(1)} (q,\omega)=
\frac{1}{N} \int_{-\infty}^{+\infty}\int_{-\infty}^{+\infty}
d \omega' d\omega_1 \frac{n(\omega_1-\omega') - n(\omega_1)}
{\omega-\omega'+ \imath \varepsilon}
$$
\begin{equation}
\times \sum_{k,\sigma}
t_{kq}^2 A^{\sigma} (k-q, \omega_1-\omega ') A^{\bar\sigma}(k,\omega_1 )
\label{g71}
\end{equation}
where $t_{kq}=  zt (\gamma_{k}- \gamma_{k-q}) $,
$n(\omega) =(e^{\beta\omega}+1)^{-1}$,
and $A^{\sigma}(k,\omega )$
is the single-electron spectral function. By using the MF approximation
for that, Eq.~(A.7), we can integrate over the frequencies in Eq.~(\ref{g71})
and obtain the following estimation:
\begin{equation}
\Pi_{11}^{(1)} (q,\omega)=
 \frac{1}{N} \sum_{k, \sigma}t_{kq}^{2}
(1-n^{\sigma})(1-n^{\bar\sigma})
 \frac {n(\varepsilon_{k-q}^{\sigma}) - n(\varepsilon_{k}^{\bar\sigma})}
 {\omega+\varepsilon_{k-q}^{\sigma} - \varepsilon_{k}^{\bar\sigma} +
 \imath\varepsilon}  \; .
\label{g71a}
\end{equation}
It has the standard form for a one-loop particle-hole contribution to
the self-energy (see, e.g.~\cite{izyum}).

The second terms in Eq.(\ref{g70})
are the same for both subsystem and coincides with the nondiagonal
elements of the self-energy matrix due to the Hund coupling
\begin{equation}
\Pi_{11(22)}^{(2)}(q,\omega)= - \Pi_{12(21)}(q,\omega)=
\Pi_{H}(q,\omega)
\end{equation}
with
\begin{eqnarray}
\Pi_{H}(q,\omega)= \left( \frac{J_H}{2S} \right) ^2
\frac{1}{N\pi^2}  \int_{-\infty}^{+\infty}\int_{-\infty}^{+\infty}
d \omega ' d\omega_1
\frac{1+N(\omega '-\omega_1)+N(\omega_1)}
{\omega-\omega '+ \imath \varepsilon}\nonumber\\
\times \sum_{k}
 \left[{\rm Im}\chi_{22}^{z}(k,\omega_1 ){\rm Im}
\chi_{11}(k-q,\omega '-\omega_1)
+{\rm Im}\chi_{22}(k,\omega_1 ){\rm Im}\chi_{11}^{z}(k-q,\omega '-\omega_1)
\right]
\label{g81a}
\end{eqnarray}
where $N(\omega) =(e^{\beta\omega}-1)^{-1}$,
$\chi_{11}^{z}$ and $\chi_{22}^{z}$ denotes the longitudinal susceptibility
of the itinerant and core spins, respectively.

Let us consider now the remaining
$\Pi_{22}^{(1)}$ term which  describes the fluctuations in
the core spin subsystem. This contribution  is due to the
Heisenberg exchange between the localized spins and is given by
\begin{eqnarray}
\Pi_{22}^{(1)}(q,\omega)=\frac{1}{N\pi^2}
  \int_{-\infty}^{+\infty}\int_{-\infty}^{+\infty} d \omega ' d\omega_1
\frac{1+N(\omega '-\omega_1)+N(\omega_1)}
{\omega-\omega '+ \imath \varepsilon}\nonumber\\
\times \sum_{k}J_{kq}^2{\rm Im}\chi_{22}^{z}(k-q,\omega_1)
{\rm Im}\chi_{22} (k, \omega'-\omega_1) \; ,
\label{g92}
\end{eqnarray}
where $J_{kq} = zJ (\gamma_{k}- \gamma_{k-q})$.

In order to evaluate the longitudinal susceptibility in
Eqs.~(\ref{g81a}),~(\ref{g92}) for
both subsystems  we will use for them the simplest one-loop approximation
(see, e.g.~\cite{izyum}).  In this approximation the imaginary part
of  $\chi_{11}^{z} (q, \omega)$ is given
as the convolution of the  single-electron GFs
$$
-\frac{1}{\pi}{\rm Im}\chi_{11}^{z} (q, \omega)=
\frac{1}{4N} \int_{-\infty}^{+\infty} d \omega '
[n(\omega'-\omega )-n(\omega ')]
$$
\begin{equation}
\times  \sum_{k,\sigma}
A^{\sigma}(k,\omega ')A^{\sigma}
 (k-q,\omega '-\omega).
\label{chi11z}
\end{equation}
The imaginary part of the core
spin susceptibility  $\chi_{22}^{z}$  can be expressed
in the linear spin-wave approximation  as
$$
-\frac{1}{\pi} {\rm Im}\chi_{22}^{z} (q, \omega)=
\frac{1}{\pi^{2} 4 S^{2} N} \int_{-\infty}^{+\infty}
d\omega' ( N(\omega'-\omega)-N(\omega') )
$$
\begin{equation}
 \sum_{k} {\rm Im} \chi_{22} (k-q, \omega' - \omega)
{\rm Im} \chi_{22} (k, \omega')  \; ,
\label{chi22z}
\end{equation}
which follows directly from the Holstein-Primakoff representation.

To study the spin wave spectrum including self-energy corrections
let us consider the static  limit for $q\rightarrow 0$. For the self-energy
matrix we can write
\begin{equation}
\lim _{q\rightarrow 0}\Pi (q,0)=
\left(
\begin{array}{cc}
 - A q^2 - d_1 & d_1\\
 d_1 & - B q^2 - d_1 \\
\end{array}
\right)
\label{g110}
\end{equation}
where
\begin{equation}
A= - \lim\limits_{q\to 0} \frac{\Pi_{11}^{(2)}(q,0)}{q^{2}},\;
B= - \lim\limits_{q\to 0}\frac{\Pi_{22}^{(2)}(q,0)}{q^{2}},\;
d_1= - \Pi^{J_{H}}(0,0).
\label{g110a}
\end{equation}
Here  the coefficient $\; A, B,$ and $ d_1 \; $ are positive
since $\Pi (q,0) < 0$ in the second order of the perturbation theory.
As it follows from Eqs. (\ref{g110}) and
(\ref{g110a}), the self-energy corrections  coming from the Hund's coupling
does not renormalize the spin stiffness   and gives input only into the gap.
Hence the spin-wave spectrum in the longwavelegth limit can be written as
\begin{eqnarray}
\omega^{(1)}_q\simeq\tilde {D_1}q^2
\nonumber \\
\omega^{(2)}_q\simeq\tilde {\Delta}
+\tilde {D_2} q^2
\label{g113}
\end{eqnarray}
where the renormalized spin stiffness and the gap are given by
\begin{eqnarray}
\tilde {D_1}=
\frac{a- b - A - B}
{12(\langle S^z \rangle +\langle \sigma ^z\rangle)}\nonumber\\
\tilde {D_2}  =
\frac{(a - A)\langle S^z \rangle^2 -(b+B)\langle \sigma^z\rangle^2}
{12 \langle S^z \rangle\langle \sigma ^z\rangle
(\langle S^z \rangle +\langle \sigma ^z\rangle)}\label{ddd}\\
\tilde {\Delta}
=\frac{(d- d_1)(\langle S^z \rangle +\langle \sigma ^z\rangle)}
{2\langle S^z\rangle\langle \sigma ^z\rangle}.
\nonumber
\end{eqnarray}

Now we consider the  damping for the acoustic  spin wave mode given by the
imaginary parts of the self-energy matrix.
For the damping induced by  particle-hole excitations we get from
Eq.~(\ref{g71a}) in the MF approximation for the single-electron GFs:
$$
\Gamma_{11}^{(1)} (q,\omega)=
- {\rm Im}\Pi_{11}^{(1)} (q,\omega+\imath \varepsilon)
$$
\begin{equation}
= \frac{\pi}{N} \sum_{k, \sigma}t_{kq}^{2}
(1-n^{\sigma})(1-n^{\bar\sigma})
[n(\varepsilon_{k-q}^{\sigma}) - n(\varepsilon_{k}^{\bar\sigma})]
\delta (\omega+\varepsilon_{k-q}^{\sigma}-\varepsilon_{k}^{\bar\sigma})
\label{p11}
\end{equation}
The contribution,  due to the finite $k-$independent
gap in single-electron  spectrum in the ferromagnetic state
 disappears in the low frequency limit,
$ \Gamma_{11}^{(1)} (q,\omega)= 0$  for $\omega < h $ (see Eq.~(A.14)).

The damping due to the antiferomagnetic exchange
interaction given by the imaginary part of the self-energy
$\Pi_{22}^{(1)} (q,\omega)$, Eq.~(\ref{g92}), gives a small contribution
proportional to $q^2$ in the longwavelength limit and can be disregarded
due to small antiferromagnetic exchange interaction $J$.

The largest contribution to the damping of spin waves is given by the
imaginary part of the self-energy due to the Hund coupling, Eq.~(\ref{g81a}):
$$
\Gamma_{H} (q,\omega)=
 - {\rm Im}\Pi_{H}(q, \omega +\imath \varepsilon) =
 \left(\frac{J_H}{2S}\right)^2
\frac{1}{\pi N}\int_{-\infty}^{+\infty} d\omega_1
 (1+N(\omega -\omega_1)+N(\omega_1))
$$
\begin{equation}
\times \sum_{k}\left[ {\rm Im}\chi_{22} (k-q, \omega - \omega_1)
{\rm Im}\chi_{11}^{z}(k, \omega_1) +
{\rm Im}\chi_{11}(k-q, \omega -\omega_1)
{\rm Im}\chi_{22}^z (k, \omega_1) \right] .
\label{g82}
\end{equation}
Here for  the imaginary parts of the spin susceptibilities
$\chi_{11}(k,\omega)$ and  $\chi_{22}(k,\omega)$
 we will use their MF values in Eqs.~(\ref{chi11}),~(\ref{chi22})
 taking into account  only the acoustic $E_q^{(1)} $ mode:
\begin{eqnarray}
-\frac{1}{\pi}{\rm Im}\chi_{11}^{MF} (q,\omega) \simeq
 2 \langle \sigma ^z\rangle
 \frac{ {\tilde\Omega}^{22}_q - E^{(1)}_q }{E^{(2)}_q-E_q^{(1)}}
\delta(\omega-E^{(1)}_q) =\Lambda^{11}_{q} \delta(\omega-E_q) \; ,
\label{c1}
\end{eqnarray}
\begin{eqnarray}
-\frac{1}{\pi}{\rm Im}\chi_{22}^{MF} (q, \omega) \simeq
2 \langle S^z\rangle
 \frac{  {\tilde\Omega}^{11}_q - E^{(1)}_q }{E^{(2)}_q-E_q^{(1)}}
\delta(\omega-E^{(1)}_q) =  \Lambda^{22}_{q} \delta(\omega-E_q) \; .
\label{c2}
\end{eqnarray}
For the longitudinal spin susceptibilities in (\ref{g82}) we will use
 Eqs.~(\ref{chi11z}), (\ref{chi22z}).

After integration over $\omega_1$ we get the following result
$$
\Gamma_{H}(q, \omega) = (e^{{\omega}/{T}}-1)\left(\frac{J_H}{2S}\right)^2
\frac{\pi}{4N^2} \sum_{k,k_1,\sigma} \Lambda^{22}_{k-q} (1-n_{\sigma})^2
$$
$$
\times  N(E_{k-q}) n(\varepsilon_{k_1}^{\sigma})
[1 - n(\varepsilon_{k_1-k}^{\sigma})]
\delta (\omega - E_{k-q} +\varepsilon_{k_{1}-k}^{\sigma}-
\varepsilon_{k_{1}}^{\sigma})
$$
$$
 +(e^{{\omega}/{T}}-1)\left(\frac{J_H}{2S}\right)^2
\frac{\pi}{4 S^2 N^2} \sum_{k,k_1}
 \Lambda^{11}_{k - q}\Lambda^{22}_{k_1}\Lambda^{22}_{k_1 - k}
$$
\begin{equation}
\times  N(E_{k-q})  N(E_{k_1})[1+N(E_{k_1-k})]
\delta (\omega- E_{k-q}^{(1)}+E_{k_{1}- k}^{(1)} - E_{k_1}) ~.
\label{g86}
\end{equation}
It describes a spin wave damping due  to its decay  into an electron-hole
pair and another spin wave (the first term) and a three spin-wave scattering
process (the second term). The latter has a standard form
for three magnon scattering
(see, (31.2.20) in \cite{ax}).
At low energy regime ($\omega\rightarrow 0$) and at the longwavelenght limit
the requirements for conservation of energy and momentum
allow only small  wave vectors and thus only small energies.
Hence we can consider $\Lambda^{11}_{k}$ and $\Lambda^{22}_{k}$   as
$k$-independent. In the limit $k \rightarrow 0$ we obtain
\begin{equation}
\Lambda^{11}_{k} \simeq \Lambda^{11}_0 = 2 \langle \sigma^z\rangle
\frac{\langle \sigma^z\rangle}{\langle S^z\rangle+\langle \sigma^z\rangle}~,~
\Lambda^{22}_{k} \simeq \Lambda^{11}_0 = 2 \langle S^z\rangle
\frac{\langle S^z\rangle} {\langle S^z\rangle+\langle \sigma^z\rangle}\; .
\label{lambda}
\end{equation}
In this approximation  from the Eq.~(\ref{g86})  follows  that
at low energies, ($\omega \ll T$), the damping
has a linear $\omega$-dependence,
$\Gamma^{J_H}(q, \omega)\sim \omega$, and  does not depend explicitly
on the wave vector $q$. In the low temperature region,
($\omega \ll T \ll \omega_0$) where $\omega_0 $ is the maximal acoustic
spin-wave frequency, estimations for $\; \Gamma_{H}(q, \omega)/\omega \;$
show that the first contribution in
Eq.~(\ref{g86}) is proportional to
$\;(T/\omega_{0})( J_{H}^{2}/ N(\epsilon_{F}) v_{F}k_{0}) \;$
where $N(\epsilon_{F})$ is the density of state at the Fermi level and
$ v_{F}$ is the Fermi velocity,  $k_{0}\simeq 2\pi/a$ and $a$ is a lattice
constant. The second term gives to
the damping the contribution proportional
to $\; (T/\omega_{0})( J_{H}/ \omega_{0})^{2} \;$.
To give  more accurate estimations
for the spin-wave damping numerical studies
should be performed which  will be considered elsewhere.

\section{Conclusions}

In the present paper we have calculated dynamical spin susceptibility
for the generalized ferromagnetic Kondo model (1) by taking into account
explicitly both the strong Hund  interaction for the itinerant $e_g$  and
localized $t_{2g}$  electrons  and AFM interaction between the  $t_{2g}$
 electrons. We consider the ferromagnetic phase and therefore neglect
a possible orbital ordering of the $e_g$ electrons.
 Strong electron correlations between  $e_g$ electrons are treated
 within the Hubbard operator technique which is important in calculation
 of the single-electron GF for the itinerant electrons.

 We have proved that even in the MFA described by the frequency matrix,
 Eq.~(19), we get the , acoustic spin-wave excitations, Eq.~(24),
 due to  coupling of the two modes with gaps for itinerant and localized
 electrons. The gapless mode should appear in the model (1) with rotation
 symmetry for spin system.
A gapless mode in the limit $J_{H} \to \infty$, considered in Ref.~\cite{fur},
was obtained only by taking into account self-energy corrections. In our
case the self energy corrections calculated in the self-consistent Born
approximation, Eqs.~(32)-(34), resulted in additional renormalization of the
stiffness of the acoustic ferromagnetic spin waves. The imaginary parts of the
self-energy gives the damping of spin waves. We have evaluated the most
important contribution due to Hund coupling in the second order, Eq.~(49),
which can be described as a three magnon scattering. The damping of acoustic
spin waves is proportional to the frequency for $\omega\ll T$, Eq.~(51), and should be small
for small wave vectors. To give numerical estimations one should solve
self-consistently the system of equations for the matrix spin susceptibility,
Eq.~(6), and the self-energy functions, Eq.~(37). Also the spectrum of single
electron excitations, given in the Appendix, should be evaluated to consider
the itinerant electron contributions to the spin waves, Eq.~(36). These problems
will be considered elsewhere.\\

{\bf Acknowledgments}\\
We wish to thank G.Jackeli for valuable discussions.
A partial financial support by the INTAS Program,
Grant No 97-0963, is acknowledged.

\section*{Appendix}
In this Appendix we evaluate the single-electron Green
function defined as
$$
G^{\sigma}(k,\omega)=\langle\langle X_k^{0\sigma}|X_k^{\sigma 0}
\rangle\rangle_{\omega}
\quad \eqno (A.1)
$$
with the corresponding spectral function
$A _{k}^{\sigma} (\omega)=
-({1}/{\pi}) {\rm Im}~G^{\sigma}(k,\omega +\imath \varepsilon)$.
In the site representation the equation of motion for $G^{\sigma}(k,\omega)$
reads as
$$
\omega\langle\langle
 X_i^{0\sigma}|X_k^{\sigma 0}\rangle\rangle_{\omega}=
\langle\left\{X_{i}^{0\sigma} ,X_{j}^{\sigma 0}\right\}\rangle+
\langle\langle\imath\dot{X_i^{0\sigma}}|X_j^{\sigma 0}\rangle\rangle _{\omega}~.
\quad \eqno (A.2)
$$
Using the   commutation relations, Eq.(\ref{m7}), we obtain
$$
\imath \dot {X_{i}^{0\sigma}}=[X_{i}^{0\sigma}, H]
=-\sum_{l\neq i}t_{il}
\left[(X_{i}^{00}+X_{i}^{{\sigma}{\sigma}})X_{l}^{0\sigma}+
X_{i}^{\bar{\sigma}\sigma}X_{l}^{0\bar{\sigma}}\right]
$$
$$-\frac{J_H}{4S}\left[\sigma S_{i}^{z}X_{i}^{0\sigma} +
S_{i}^{\bar{\sigma}}X_{i}^{0\bar{\sigma}}\right]~.
\quad \eqno (A.3)
$$
The next step is to define the irreducible part $Z_{i}^{0\sigma}$ of
the current operator $\imath \dot{X_{i}^{0\sigma}}$
by
$$\imath \dot{X_{i}^{0\sigma}}=\sum_{l}\varepsilon_{il}^{\sigma}
X_{l}^{0\sigma}+
Z_{i}^{0\sigma}~,~
\langle \{Z_{i}^{0\sigma},X_{j}^{\sigma 0} \} \rangle=0~.
\quad \eqno (A.4)
$$
The definition gives for the frequency matrix
$$
\varepsilon_{ij}^{\sigma}=
\langle\{\imath{\dot X}_{i}^{0\sigma},X_{j}^{\sigma 0}\}\rangle
/(1-n^{\bar{\sigma}}) ,
\quad \eqno (A.5)
$$
where by using the completeness relation, Eq.(\ref{m4}), we write
$\;\langle\left\{X_{i}^{0\sigma}, X_{j}^{\sigma 0}\right\}\rangle=
\delta_{i,j} (1-n^{\bar{\sigma}}) \; $
with $n^{\bar{\sigma}} = \langle n^{\bar{\sigma}}_{i} \rangle $.
By using equation of motion (A.3) we get
$$
\langle\{\imath{\dot X}_{i}^{0\sigma},X_{j}^{\sigma 0}\}\rangle =
-t_{ij}\langle (1-n^{\bar{\sigma}}_i)(1-n^{\bar{\sigma}}_j)\rangle
-t_{ij}
\langle X^{\bar{\sigma}\sigma}_{i} X_{j}^{\sigma\bar{\sigma}}\rangle
$$
$$+\sum_{l}t_{il}\langle X^{\bar{\sigma}0}_{i} X_{l}^{0\bar{\sigma}}\rangle
-\frac{J_H}{4S}\sigma\langle S_i^z \rangle (1-n^{\bar{\sigma}})\delta_{ij}
 - \frac{J_H}{4S}\langle S^{\bar{\sigma}}_{i} X_{i}^{\sigma\bar{\sigma}}\rangle
\delta_{ij}~.
\quad \eqno (A.6)
$$
In the present paper we will not include the self-energy correction
coming from $Z_{i}^{0\sigma}$  term (A.4) and
treat the single-electron GF within the linear, MF type approximation.
That results in the following form of the  single-electron GF
$$
G^{\sigma}(k, \omega)=\frac{1-n^{\bar {\sigma}}}
{\omega-\varepsilon_{ k}^{\sigma} }.
\quad \eqno (A.7)$$
By introducing the nearest neighbor charge-spin correlation function
$$N_{1,\sigma}^{cs}=\frac{1}{N}\sum_{k}\gamma_{k}\left[
\langle X_k^{\bar{\sigma} \bar{\sigma}} X_{-k}^{\bar{\sigma} \bar{\sigma}}\rangle+
\langle X_k^{\bar{\sigma} \sigma} X_k^{\sigma \bar{\sigma}}\rangle\right]
$$
$$
=\langle X_{i}^{\bar{\sigma} \bar{\sigma}}
X_{i+a}^{\bar{\sigma} \bar{\sigma}}\rangle+
\langle X_i^{\bar{\sigma} \sigma} X_{i+a}^{\sigma\bar{\sigma}}\rangle
\quad \eqno (A.8) $$
we can rewrite the frequency matrix in the form
$$
\varepsilon_{ij}^{\sigma}=
\epsilon^{\sigma}\delta_{ij}+\epsilon_{ij}^{\sigma}~.
\quad \eqno (A.9)
$$
 where
$$ \epsilon_{ij}^{\sigma}
 =t_{ij}[(1-2n^{\bar{\sigma}}) + N_{1,\sigma}^{cs}]
   /(1-n^{\bar{\sigma}})
\quad \eqno (A.10) $$
 is the one particle spectrum in the linear approximation, and
$$ \epsilon^{\sigma}= -\frac{J_H}{4S}\left[
 \sigma\langle S^{z}\rangle+
 \frac{\langle S_{i}^{\bar{\sigma}}X_{i}^{\sigma\bar{\sigma}}\rangle }{
 1-n^{\bar{\sigma}}}
 \right]+
 \frac{z t n_{1}^{\bar{\sigma}}}{1-n^{\bar{\sigma}}} ,
\quad \eqno (A.11)
$$
 is the spin- dependent energy shift of the spectrum.
In the momentum space the spectrum is
$$
\varepsilon_{k}^{\sigma} =\sum_{R_{ij}}e^{-\imath k R_{ij}}
\varepsilon_{ij}^{\sigma}=\epsilon^{\sigma}-zt^{\sigma}_{eff}\gamma_{k} ,
\quad \eqno (A.12)
$$
where
$$
t^{\sigma}_{eff}=t
[(1-2n^{\bar{\sigma}}) + N_{1,\sigma}^{cs}]
  / (1-n^{\bar{\sigma}} )  ,
\quad \eqno (A.13)
$$
is an effective bandwidth that is  narrowed  and spin dependent
due to the spin and charge  correlations. In the MFA  in the ferromagnetic
state we have a spin gap in the single-electron spectrum:
$$
\varepsilon_{k}^{\bar \sigma} - \varepsilon_{k}^{\sigma}  \simeq
 \sigma (J_{H}/2S ) \langle S^{z}\rangle = \sigma h .
\quad \eqno (A.14)
$$


\begin{thebibliography}{200}

\bibitem{ram} A.P.Ramirez, J. Phys. Cond. Matter {\bf 9}, 8171, (1997).

\bibitem{mil} A.J.Millis et al.,
Phys. Rev. B {\bf 54}, 5389 and 5405, (1996).

\bibitem{khom} D.I.Khomskii, G.A.Sawatzky,
Sol.St.Comm.{\bf 102}, 87, (1997).

\bibitem{good}  J.B.Goodenough, {\it Magnetism and Chemical Bonds},
Intersc. publ., NY, (1963).

\bibitem{zener} C. Zener, Phys.Rev. {\bf 82}, 403, (1951)

\bibitem{and} P.W.Anderson, H.Hasegawa, Phys.Rev. {\bf 100}, 675, (1955).

\bibitem{degen} P.G. de Gennes, Phys.Rev. {\bf 118}, 141, (1960).

\bibitem{shif} P.Schieffer et al., Phys. Rev. Lett. {\bf 75}, 3396, (1995).

\bibitem{ino}  J.Inoue, S.Maekawa, Phys. Rev. Lett. {\bf 74}, 3407, (1995).

\bibitem{ish}  S. Ishihara, J.Inoue, S.Maekawa,
Phys.Rev.B {\bf 55}, 8280, (1997).

\bibitem{balag}   A.M.Balagurov et al., JETP Lett., to be published.

\bibitem{per} T.G.Perring et al., Phys. Rev. Lett. {\bf 78}, 3197, (1997).

\bibitem{lynn} J.W.Lynn et al., Phys. Rev. Lett. {\bf 76}, 4046, (1996).

\bibitem{fur} N.Furukawa, J. Phys. Soc. Jap. {\bf 63}, 3214, (1994).

\bibitem{fei} L.F.Feiner and A.M.Oles, Phys. Rev. B {\bf 56}, in press (1998).

\bibitem{sap} S.Satpathy, Z.S.Popovic, F.R.Vukajlovic,
Phys.Rev.Lett. {\bf 76}, 960, (1996).

\bibitem{got} W. Gotse and P.Wolfle, J. Low  Temp. Phys. {\bf 5}, 575 (1971).

\bibitem{zub} D. N. Zubarev, Usp. Fiz. Nauk. {\bf 71},
	      71, (1960).
\bibitem{tse} Yu. A. Tserkovnikov, Theor. Math. Fiz. {\bf 49},
219, (1981); {\bf 52},  147, (1982).
\bibitem{izyum}  Yu.A. Izyumov, M.I. Katsnelson, and Yu.N. Skriabin,
Magnetism of itinerant electrons, (Moscow, Nauka) (1994).
\bibitem{ax} A.I.Axiezer, V.G.Bariyaxtar, S.V.Peletminskii,
{ Spin waves}, (Moscow, Nauka) (1967).

%\bibitem{dagot} J. Riera, K. Halberg, E. Dagotto, Phys. Rev. Lett. {\bf 79},
%713, (1997).

\end{thebibliography}
\end{document}